\documentclass[aps,twocolumn,
superscriptaddress,
footinbib,
prb]{revtex4-2}

\newcommand{\ceqn}[1]{Eq.~\ref{#1}}

\newcommand{\cref}[1]{Ref.~\cite{#1}}

\newcommand{\cact}[1]{\hyperref[#1]{#1}}
\newcommand{\ignore}[1]{}

\usepackage{amsmath,amssymb,bm}
\usepackage{graphicx}
\usepackage[usenames,dvipsnames]{color}
\usepackage{hyperref}
\usepackage{natbib}
\usepackage{xcolor}
\usepackage{braket}
\usepackage{tikz}
\usepackage{dsfont}
\usetikzlibrary{shapes}
\usetikzlibrary{calc}

\usepackage{MnSymbol} 

\usepackage[utf8]{inputenc}

\begin{document}

\title{Geometric inflation of deviations challenges neural quantum states in dynamics of quantum Ising models}

\date{\today}
\author{Wladislaw Krinitsin}~\thanks{These authors contributed equally to this work.} 
\email{w.krinitsin@fz-juelich.de}
	\affiliation{Forschungszentrum J\"ulich GmbH, Peter Gr\"unberg Institute,
Quantum Control (PGI-8), 52425 J\"ulich, Germany}
    \affiliation{University of Regensburg, 93053 Regensburg, Germany}
\author{Jonas B. Rigo}~\thanks{These authors contributed equally to this work.} 
\email{jonas.rigo@ur.de}
	\affiliation{Forschungszentrum J\"ulich GmbH, Peter Gr\"unberg Institute,
Quantum Control (PGI-8), 52425 J\"ulich, Germany}
    \affiliation{University of Regensburg, 93053 Regensburg, Germany}
\author{Mohammad Abedi}
\email{m.abedi@fz-juelich.de}
	\affiliation{Forschungszentrum J\"ulich GmbH, Peter Gr\"unberg Institute,
Quantum Control (PGI-8), 52425 J\"ulich, Germany}
    \affiliation{University of Regensburg, 93053 Regensburg, Germany}
\author{Markus Schmitt}
\email{markus.schmitt@ur.de}
	\affiliation{Forschungszentrum J\"ulich GmbH, Peter Gr\"unberg Institute,
Quantum Control (PGI-8), 52425 J\"ulich, Germany}
    \affiliation{University of Regensburg, 93053 Regensburg, Germany}

\begin{abstract}
Neural quantum states (NQS) have emerged as a powerful framework for simulating non-equilibrium dynamics in strongly correlated quantum systems, offering scalable variational representations of highly entangled states.
Yet, accurate NQS simulations have been found to be surprisingly challenging in some physical regimes of limited complexity.
Here, we address paradigmatic quench dynamics of a one-dimensional quantum Ising model as a controlled benchmark.
Through supervised state reconstruction we establish substantially tighter empirical upper bounds on the required parameter count than previous estimates, ruling out representational limitations as the key obstruction. 
Instead, we uncover a geometric inflation of small deviations as a hitherto overlooked challenge for accurate solutions of the infinitesimal time-dependent variational principle (TDVP): the dynamical rotation of the kernel of the quantum geometric tensor (QGT) can suddenly lend physical significance to previously irrelevant parameter deviations.
The stability of matrix product state solutions of the same TDVP suggests that the non-linearity of the neural network ansatz is the origin of the sensitivity.
These results identify QGT-null-space rotation as a geometric diagnostic of sensitive NQS dynamics and as a concrete target for improving TDVP algorithms.
\end{abstract}

\maketitle

\section{Introduction}

Non-equilibrium quantum many-body physics connects fundamental questions of statistical mechanics with the control of quantum matter and emerging quantum technologies.
Rapid progress in quantum simulators, quantum computers, and ultrafast probes has exposed phenomena such as discrete time crystals~\cite{Zhang_2017,Choi_2017,Moon_2025}, light-induced superconductivity~\cite{Fausti_2011,Budden_2021}, many-body localization~\cite{Schreiber_2015,Choi_2016}, and long-lived prethermal regimes~\cite{Birnkammer_2022}.
Interpreting these experiments requires reliable numerical access to quantum dynamics, yet the exponential growth of Hilbert space and the rapid production of entanglement severely restrict exact methods and tensor-network calculations, particularly in two and higher spatial dimensions~\cite{schollwoeck2011density,paeckel2019time}.
This leaves a central computational challenge: finding compressed representations that remain both expressive and dynamically stable as correlations spread.
Neural quantum states (NQS) pursue this goal by using artificial neural networks as variational representations of many-body wave functions~\cite{carleo}.
Their promise rests on the same properties that make neural networks effective high-dimensional function approximators: the ability to capture recurring structures and complex correlations efficiently, together with universal approximation as network capacity increases~\cite{hornik1989multilayer,cybenko1989approximation}.
Moreover, the known bounds for NQS sizes required to encode given amounts of entanglement leave room for highly efficient representation of strongly entangled states \cite{levine_quantum_2019,paul2025bound}.
This combination has enabled increasingly accurate ground-state calculations for frustrated two-dimensional magnets and interacting fermions, including state-of-the-art results for the square-lattice $J_1$--$J_2$ and Hubbard models~\cite{chen2024empowering,chen2025scalable,roth2025superconductivity,viteritti2026beyond,gu2026pareto,viteritti2026approaching}.
NQS-based dynamics has reached large systems in one, two, and recently three dimensions and has enabled calculations of quench dynamics, spectral functions, and phase transition dynamics beyond the regimes accessible to exact methods~\cite{schmitt2020quantum,schmitt_quantum_2022,MendesSantos2023,nys2024ab,naik2026realtime}.
Nevertheless, progress in dynamics has not yet acquired the same systematic reliability as ground-state optimization: reachable timescales have remained limited in some cases without clear reasons and the resulting accuracy can depend sensitively on details such as the chosen architecture, sampling procedure, or required regularization for stable time propagation~\cite{czischek2018quenches,hofmann2022RoleStochasticNoise,donatella2023dynamics,sinibaldi2023unbiasing,misery2025looking,king2025aBeyondclassicalComputation,vovrosh2025SimulatingDynamics,xue2026direct,haghshenas2026DigitalQuantumMagnetism,krinitsin_comment_2026}.

\begin{figure*}[ht]
  \centering
  \includegraphics[width=\linewidth]{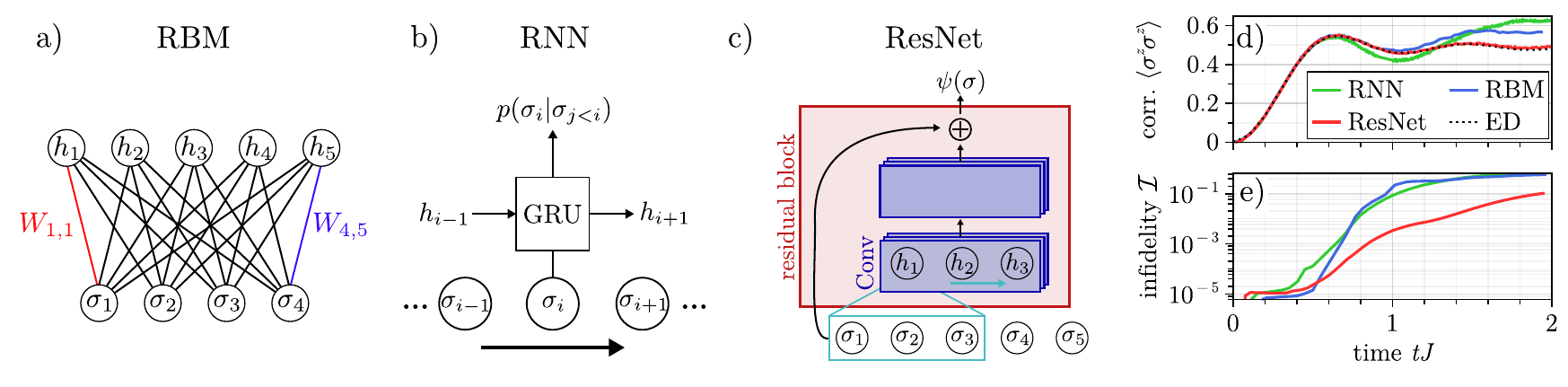}
  \caption{Schematic representations of different neural network architectures: a) the RBM, b) the RNN, and c) the ResNet. The latter consists of residual blocks that can be chained to increase the network depth, which is set to $\mathrm{depth}=2$ in this work. Panels d) and e) show that NQS encounter an obstruction early in the evolution of an $x$-polarized state quenched to the critical TFIM in one dimension for $L=20$ spins. The analysis uses the three aforementioned architectures with $N_P=14480$ (RBM), $N_P=8434$ (RNN), and $N_P=12200$ (ResNet) parameters. d) While the $\langle \sigma^z\sigma^z \rangle$ correlation agrees with the exact reference result up to later times---especially for the ResNet---e) the infidelities between the variational and exact time-evolved states start to grow at around $tJ\approx 0.5$ or earlier, signaling a steadily increasing deviation.} \label{fig:crit_tfim_failure}
\end{figure*}

Several distinct explanations for the observed fragility of time evolution algorithms have been put forward.
In its infinitesimal formulation, the time-dependent variational principle (TDVP) employed for time-propagation requires stochastic estimation of a non-linear differential equation \cite{Carleo2012,carleo,schmitt_simulating_2025}. The inevitable sampling noise has been identified as a source of instability, requiring careful regularization \cite{schmitt2020quantum,donatella2023dynamics}. Moreover, zeros of the wave function can bias conventional Monte Carlo estimators \cite{sinibaldi2023unbiasing}, motivating the construction of alternative estimators \cite{krinitsin2026} or global optimization approaches to solve the TDVP~\cite{sinibaldi2023unbiasing, gravinaNeuralProjectedQuantum2025a}.
From the perspective of representational power, Lin and Pollmann reported a rapid growth of the parameter count required to maintain a fixed accuracy during time evolution~\cite{lin2022scaling}, and Dash \textit{et al.}~\cite{dash2025efficiency} connected the practical representation power of NQS to the rank of the quantum geometric tensor (QGT) describing the local geometry of the NQS manifold.
These mechanisms can coexist, interact, and reinforce one another, rendering the analysis and identification of ultimate bottlenecks challenging.
The underlying reasons for the observed limitations have thus often remained opaque---especially in cases, where accurate NQS simulations have been restricted to surprisingly short times.

In this work, we address a paradigmatic example, where NQS simulations become inaccurate at early times: quenches from a product state to the critical region in one-dimensional quantum Ising models.
This case has been identified early-on as particularly challenging by Czischek et al.~\cite{czischek2018quenches}: Even for system sizes amenable to exact numerical simulation, the NQS simulations using a Restricted Boltzmann Machine (RBM) ansatz could not be converged beyond the time needed for building up nearest-neighbor correlations, which matrix product states capture comfortably with modest bond dimensions.
The finding of Ref.~\cite{czischek2018quenches} is qualitatively reproduced in Fig.~\ref{fig:crit_tfim_failure}d), where a clear deviation of the RBM simulation becomes apparent in the displayed nearest-neighbor correlation function soon after build-up.
We find that an alternative choice of the architecture improves the accuracy of the observable, but as shown in Fig.~\ref{fig:crit_tfim_failure}d) a distinct increase of the infidelity with the exactly time-evolved wave function remains unavoidable up to the largest tractable neural network sizes.
The critical quantum Ising model therefore constitutes an ideal subject to investigate an inexplicably adamant bottleneck for NQS time evolution.

In the following, we present a systematic analysis ruling out the known obstacles that may limit the reliably reachable times. Instead, we identify a new culprit that has so far received little attention in the context of NQS dynamics: sensitivity to small perturbations.
We observe chaotic divergence of initially close-by trajectories in parameter space---as to be expected when integrating the non-linear TDVP equation.
This divergence, however, remains inconsequential for the represented physical state until the peculiar time, where the above-mentioned inaccuracy arises.
We uncover that this behavior has a geometric origin: the kernel of the QGT, indicating redundancies of the quantum state parameterization, changes its orientation during the dynamics.
Thereby, parameter perturbations that are weakly relevant at one time can acquire physical significance as the state evolves.
Notably, this geometric inflation of deviations appears to be a consequence of the non-linear neural network ansatz. 
The QGT kernel orientation of an MPS evolved with the same TDVP---also a non-linear dynamics in parameter space---remains almost constant over time and parameter perturbations are not affecting the represented state.
Hence, we pin-point an intricate interplay between classical chaos and geometry of the NQS manifold as a central challenge for non-linear neural network modeling of quantum dynamics.

The remainder of this paper is organized as follows: we start by introducing and motivating the main point of investigation for this work --- the quench from a product state to the critical point of the transverse field Ising model (TFIM) --- followed by a summary of the main numerical method, the TDVP for NQS. 
Section \ref{sec:expressivity} provides a comprehensive analysis of the expressive power of various ansätze in representing the evolved states at different points in time. 
Based on that, the continuity of the found solution paths is analyzed in Section \ref{sec:continuity}.
Finally, section \ref{sec:optimization} investigates optimization-related questions, in particular the sensitivity of the TDVP equation to perturbations and its geometric interpretation.
We close this work by providing a summary and discussion of the main findings in Section \ref{sec:conclusion}.

\section{Problem setup} \label{sec:problem_setup}

\subsection{Quench dynamics}
Current algorithms for time-dependent NQS can face prohibitive obstructions, typically resulting in unpredictable deviations of measured observables from reference results. 
This behavior often seems arbitrary, with no apparent ties to the underlying architecture, choice of hyperparameters and model.
Surprisingly, this phenomenology already appears in one of the arguably simplest setups --- the dynamics after a quench to the TFIM --- which constitutes the main, exemplary test case used throughout this work.
On a one-dimensional chain of length $L$ with open boundary conditions, its Hamiltonian is given by
\begin{equation}
    H_{\mathrm{TFIM}} = -J\sum_{j=1}^{L-1} \hat{\sigma}^z_j \hat{\sigma}^z_{j+1}
    - g\sum_{j=1}^{L} \hat{\sigma}^x_j,
    \label{eq:ham_tfim}
\end{equation}
Here, $J>0$ denotes the ferromagnetic coupling. 
Starting from an $x$-polarized product state, $\ket{\psi(0)}=\ket{+}^{\otimes L}$, the system is strongly quenched to the quantum critical point $g/J=1$. 
While the critical point is typically one of the most challenging to simulate, TDVP evolution with NQS has been observed to become inaccurate already very early in the dynamics~\cite{czischek2018quenches}, at which point the state exhibits at most nearest-neighbor correlations.
This is exemplified in Fig.~\ref{fig:crit_tfim_failure}~d) and e) for the case of $L=20$, using the residual network (ResNet) with $N_P=12200$ parameters: while a simple observable such as the nearest-neighbor correlation can remain accurate, the infidelity with respect to the exact reference state grows rapidly from around $tJ \simeq0.5$. 

Strikingly, other variational ansätze such as tensor networks, in particular Matrix Product States (MPS)~\cite{white1992density,schollwoeck2011density,cirac2021matrix}, are able to reach times well beyond that failure point using only moderate computational resources, to be discussed in more detail in Sec.~\ref{sec:optimization}, suggesting that the dynamics is not "hard" per se. 
As such, the presented quench constitutes a perfect testbed to investigate the capabilities of time evolution with NQS: "simple" enough to be evaluated exactly, providing a reproducible baseline for the analysis, while still exhibiting all the necessary features from the broad class of failure cases. 

In the following, we review the TDVP principle for NQS and then investigate both expressivity and optimization-related issues using Eq.~\eqref{eq:ham_tfim} as an example.

\subsection{TDVP with NQS}

The time-dependent Schrödinger equation
\begin{align}
    i\frac{d}{dt}\ket{\psi}= H\ket{\psi}
\end{align}
governs the dynamics of the quantum state given the system Hamiltonian $H$. Considering some variational encoding $\ket{\psi_\theta}$ of the wave function with parameters $\theta$, the state will be restricted to the corresponding variational manifold. As such, an approximate solution to the Schrödinger equation can be found by using the time-dependent variational principle (TDVP).
In this work, we consider different neural networks as variational ansätze, in particular the restricted Boltzmann machine (RBM), recurrent neural network (RNN), and residual network (ResNet). Graphical representations are shown in Fig.~\ref{fig:crit_tfim_failure}~a)--c), and technical details regarding the parameter scaling are provided in App.~\ref{app:networks}. 
Generally speaking, given a set of variational parameters $\theta$ and a spin configuration $\sigma$, the neural network constitutes a mapping of the form
\begin{equation}
    (\theta,\sigma) \rightarrow \psi_\theta(\sigma),
\end{equation}
which is achieved through a combination of linear layers and nonlinear functions.

Employing the stationary action principle and assuming that the solution $\theta(t)$ traces a continuous path, one can formulate an infinitesimal approximation of the Schrödinger equation projected onto the tangent space of the variational manifold, leading to a first-order differential equation for the evolution,
\begin{align}
    S_{k,k'}\dot\theta_{k'}=F_k\,. \label{eq:vmc_param_update}
\end{align}
For a real parametrization, the stationary-action equation involves $\operatorname{Im}\mathcal{G}$. To avoid dealing with a anti-symmetric matrix, we multiply this equation by $i$ and define the Hermitian S-matrix $S_{k,k'}=i\operatorname{Im}[\mathcal{G}_{k,k'}]$, where the QGT is
\begin{align}
    \mathcal{G}_{k,k'}=\frac{\braket{\partial_k\psi_\theta|\partial_{k'}\psi_\theta}}{\langle\psi_\theta|\psi_\theta\rangle}-\frac{\braket{\partial_k\psi_\theta|\psi_\theta}}{\langle\psi_\theta|\psi_\theta\rangle}\frac{\braket{\psi_\theta|\partial_{k'}\psi_\theta}}{\langle\psi_\theta|\psi_\theta\rangle}
    \label{eq:Q}
\end{align}
which encodes the local structure of the manifold at $\theta$, together with the correspondingly transformed force vector $F_k=i\operatorname{Im}[-i\mathcal{F}_k]$, where
\begin{align}
    \mathcal{F}_k=
    \frac{\braket{\partial_k\psi_\theta| H|\psi_\theta}}{\langle\psi_\theta|\psi_\theta\rangle}-
    \frac{\braket{\partial_k\psi_\theta|\psi_\theta}}{\langle\psi_\theta|\psi_\theta\rangle}
    \frac{\braket{\psi_\theta| H|\psi_\theta}}{\langle\psi_\theta|\psi_\theta\rangle}\,.\label{eq:F}
\end{align} 
The advantage of employing the stationary action principle in contrast to other variational methods lies in its property of conserving the energy during time evolution.

The quantities~\eqref{eq:Q} and~\eqref{eq:F} involve sums over the full Hilbert space, which become prohibitively expensive for large system sizes. Typically, these issues are circumvented by employing Monte Carlo sampling based on the Born distribution $p(\sigma)=|\psi_\theta(\sigma)|^2$, with $\sigma$ denoting a spin configuration. 
As has been pointed out in Ref.~\cite{sinibaldi2023unbiasing}, this prescription leads to a biased estimation of~\eqref{eq:Q} and~\eqref{eq:F}, which is especially detrimental for the time evolution. 
In recent work~\cite{krinitsin2026}, we addressed this issue by utilizing self-normalized importance sampling based on a cutoff-based probability distribution, which significantly reduces the estimation bias by encouraging exploration of the full state space.
Since the focus here lies on the expressivity and optimization of NQS, all quantities are evaluated using full sums.

After evaluating Eqs.~\eqref{eq:Q} and~\eqref{eq:F}, the differential equation~\eqref{eq:vmc_param_update} is solved using an adaptive stepper of second order, necessitating the inversion of the $S$-matrix at each step. 
To that end, we perform the eigendecomposition $S_{i,k}=\sum_j V_{i,j}\mu_j V^\dagger_{j,k}$, with $\mu_j$ and $V_{i,j}$ denoting the eigenvalues and eigenvectors, respectively. 
Generically, the S-matrix is highly ill-conditioned because of redundancies in the wave-function representation, which manifest numerically as nearly vanishing eigenvalues.
The inversion of such an ill-conditioned matrix is typically done using regularization techniques, for which in the following we present the approach used throughout this work.
Instead of using a hard cutoff, i.e. discarding all eigenvalues below some threshold, we employ a soft cutoff regularization procedure ~\cite{schmitt2020quantum} with
\begin{equation}
   \big( \mu_i\big)_\mathrm{reg}^{-1} = \left[\mu_i\left( 1+\left(\frac{\epsilon}{|{\mu_i/{\mu_1}|}}\right)^6\right) \right]^{-1}, \label{eq:lambda_inv}
\end{equation}
with the cutoff parameter $\epsilon$.
This ensures a smooth and robust inversion of the S-matrix, leading to
\begin{equation}
   \big(S_\mathrm{reg}^{-1}\big)_{i,k}
   = \sum_j V_{i,j}\big(\mu_j\big)_\mathrm{reg}^{-1}V^\dagger_{j,k}. \label{eq:S_inv}
\end{equation}
The choice of the cutoff hyperparameter $\epsilon$ will be discussed in further detail in App.~\ref{app:hyperparameter}.
For later convenience, we define the approximate nullspace or kernel of the QGT 
\begin{equation}
\ker_\alpha(\mathcal{G})=\{w_i\ |\ \mathcal{G}\, w_i=\lambda_i\, w_i \ \ \text{and} \ \ |\lambda_i/\max_j \lambda_j|<\alpha\} \label{eq:nullspace}
\end{equation} 
as the subspace spanned by eigenvectors whose eigenvalues lie below a threshold $\alpha$. These directions are approximately flat on the variational manifold. 
Parameter displacements along these directions affect the underlying physical state only weakly to leading order, i.e. they locally constitute physically weakly relevant degrees of freedom.
Conversely, the active subspace of the QGT
\begin{equation}
\mathrm{im}_\alpha(\mathcal{G})=\{w_i\ |\ \mathcal{G}\, w_i=\lambda_i\, w_i \ \ \text{and} \ \ |\lambda_i/\max_j \lambda_j|\geq\alpha\} \label{eq:activespace}
\end{equation} 
is defined as its complement. 

In the following, we analyze these quenches from three complementary perspectives: expressivity of the ansatz, continuity of accurate solutions in parameter space, and sensitivity of the TDVP flow to perturbations.

\section{Expressivity} \label{sec:expressivity}
When the time evolution of a compressed representation of a quantum state, such as an NQS, deviates from the true trajectory as shown in Fig.~\ref{fig:crit_tfim_failure}~e), the expressivity of the compressed representation is a credible culprit.

NQS can represent a reference quantum state in compressed form, but there is no direct decomposition algorithm that produces an NQS with prescribed fidelity. 
Instead, the representation has to be found by non-convex optimization, which highly depends on the choice of hyperparameters and random initializations.
This contrasts with tensor-network states, where a wave function can be decomposed and truncated systematically, yielding a compressed representation with controlled loss of precision.
Therefore for NQS, asking whether an evolution deviates from the exact trajectory because the network is not expressive enough, cannot be answered directly but only indirectly through empirically derived upper bounds on the precision that the chosen NQS can achieve for a specified target state.

The goal of this section is to find a set of variational parameters $\theta_t$ which minimize the distance between the variational state $\ket{\psi_{\theta_t}}$ and the reference state $\ket{\chi_t}$ --- obtained through exact diagonalization --- for each point in time $t$ on a specified time grid with spacing $\delta t$. 

While many valid distance measures exist, in this work we consider the infidelity loss
\begin{equation}
\mathcal{I}_t = 1-\frac{|\braket{\psi_{\theta_t}|\chi_t}|^2} {\braket{\psi_{\theta_t}|\psi_{\theta_t}}\braket{\chi_t|\chi_t}}\,, \label{eq:infidelity}
\end{equation}
which has been successfully applied in the context of projected time evolution methods~\cite{sinibaldi2023unbiasing,gravinaNeuralProjectedQuantum2025a}.
The infidelity loss has two key properties: first of all, it can be expressed as a Monte-Carlo expectation over a local estimator, allowing for its cheap estimation. 
Secondly, by using control variates, the signal to noise ratio of the estimator can be stabilized throughout the whole infidelity range. 
More details on the specifics of the algorithm can be found in App.~\ref{app:sup_learning}
\begin{figure}[t]
  \centering
  \includegraphics[width=\linewidth]{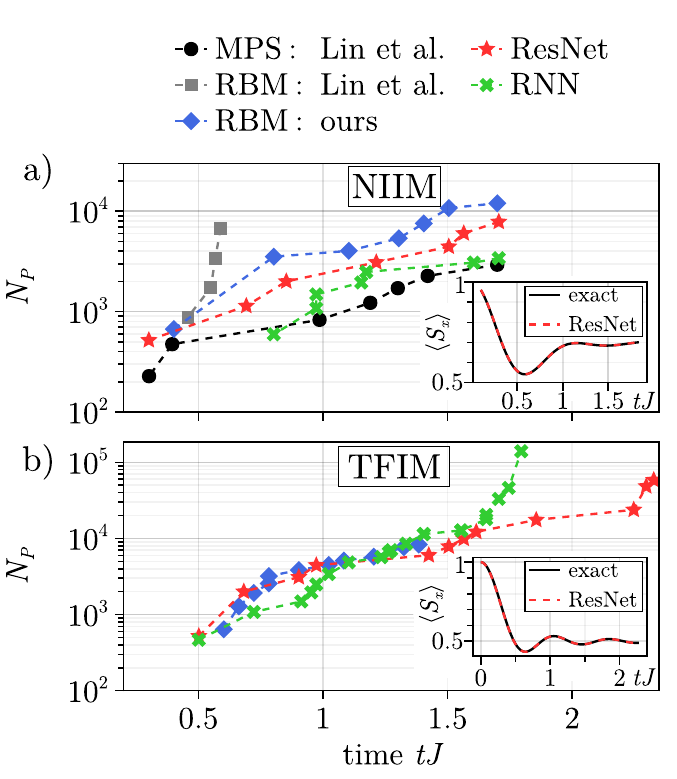}
  \caption{Minimal parameter count required to reach the target supervised accuracy $\mathcal{I}=10^{-3}$ using the RBM, RNN, and ResNet architectures during the quench dynamics of a) the NIIM, including the MPS and RBM data of Ref.~\cite{lin2022scaling}, and b) the critical TFIM. The insets compare the exact expectation value $\langle S_x\rangle$ with that produced by a ResNet with $N_P=27300$ for the NIIM and $N_P=69600$ for the TFIM.}
  \label{fig:expressivity_summary}
\end{figure}

Ultimately, this procedure yields a compressed NQS representation of the target state, with the achieved precision quantified by the final infidelity.
In order to investigate the expressivity on a whole time grid, two different approaches can be taken: either randomly initializing the network parameters at each point in time, or starting the infidelity minimization at time $t+\delta t$ using the optimal parameters found at the previous step $t$. 
Here we opt for the second option, which has the advantage of being computationally more efficient, while also opening the way to investigate the continuity of the found parameter path; to be discussed in more detail in the next Sec~\ref{sec:continuity}.
When a network, initialized with a certain number of parameters $N_P$, is not able to reach the specified target infidelity $\mathcal{I}=10^{-3}$ at time $t$, it is re-initialized at $Jt=0.1 ~ (0.5)$ for the NIIM (TFIM) model with a larger parameter count and re-optimized throughout the time grid, allowing us to map out the number of parameters needed to correctly capture the time evolution up to a certain time $t$.
The explicit architectures and the width parameters used to increase their size are summarized in App.~\ref{app:networks}.

Note that in practice, we found the optimization procedure to be reliable only for sufficiently small time steps $\delta t$.
This is due the fact that the overlap between many-body states, separated by a fixed perturbation, typically decreases exponentially with system size, requiring smaller time steps.
Thus, for larger systems, the requirement $\delta t \to 0$ may become computationally prohibitive for this form of state compression.

In order to connect the expressivity analysis to previous benchmarks, we start by considering the non-integrable Ising model (NIIM) used in Ref.~\cite{lin2022scaling},
\begin{align}
H_{\mathrm{NIIM}} = -J \bigg[
 & \sum_{j=1}^{L-1} \left( \hat{\sigma}_j^z \hat{\sigma}_{j+1}^z
+ \kappa\, \hat{\sigma}_j^x \hat{\sigma}_{j+1}^x \right) \nonumber \\
+ & \sum_{j=1}^{L} \left( g \hat{\sigma}_j^x + h \hat{\sigma}_j^z \right)
\bigg], \label{eq:ham_niim}
\end{align}
with open boundary conditions and parameters $g/J=1$, $h/J=0$, and $\kappa/J=0.25$. 
We extract the parameter count during the time evolution using the presented approach for different network architectures, in particular the RBM, RNN and ResNet, results for which are summarized in Fig.~\ref{fig:expressivity_summary}~a). 
By including the RBM as a common point of reference, our results can be directly compared to Ref.~\cite{lin2022scaling}.
Our optimized RBM follows briefly the same increasing trend as the RBM data of Ref.~\cite{lin2022scaling}, but reaches later times with substantially fewer parameters. 
The ResNet and RNN remain in the same broad range as the MPS reference over the shown time window. 
Thus, our supervised optimization protocol yields substantially lower empirical upper bounds on the parameter counts required to reach the target accuracy than those reported in Ref.~\cite{lin2022scaling}. 
The very steep scaling found by Lin and Pollmann is therefore not an unavoidable expressivity barrier of the tested NQS architectures, but can most likely be traced back to optimization specific limitations.
This analysis is repeated for the quench to the critical TFIM~\eqref{eq:ham_tfim}, see Fig.~\ref{fig:expressivity_summary}~b). 
All architectures show a rapid growth of the required parameter count, but accurate supervised representations can still be found up to $tJ \simeq2.25$ with ResNets, using parameter numbers of the order of $10^4$-$10^5$.
The insets of Fig.~\ref{fig:expressivity_summary}~a) and b) verify our optimization procedure on the level of observables, confirming that the transverse magnetization obtained from the supervised ResNet corresponds to the exact result throughout the time interval.

In view of recent work discussing connections between Walsh complexity and the expressive power of neural quantum states \cite{cortesBasisDependenceNeural2026,doschl2025importance,wang2025expressibility},
we investigate the ResNet in more detail at a representative failure time of $tJ=1$.
We start by first considering a complementary perspective provided by the cumulant, or coupled-cluster expansion, as was discussed by Cortes et al.~\cite{cortesBasisDependenceNeural2026} in the context of NQS. 
For a fixed computational basis, the coefficients of the expansion of $\log \psi(\sigma)$ can be ordered by magnitude, and a truncated state can be reconstructed from the $N_P$ largest contributions. 
For TFIM ground states, Ref.~\cite{cortesBasisDependenceNeural2026} found a power-law decay of the infidelity under this truncation, comparable to the scaling obtained with an RBM. 
This suggests that, for those states and bases, the RBM ansatz is well aligned with the dominant cumulants of the wave function.

\begin{figure}[h!]
  \centering
  \includegraphics[width=\linewidth]{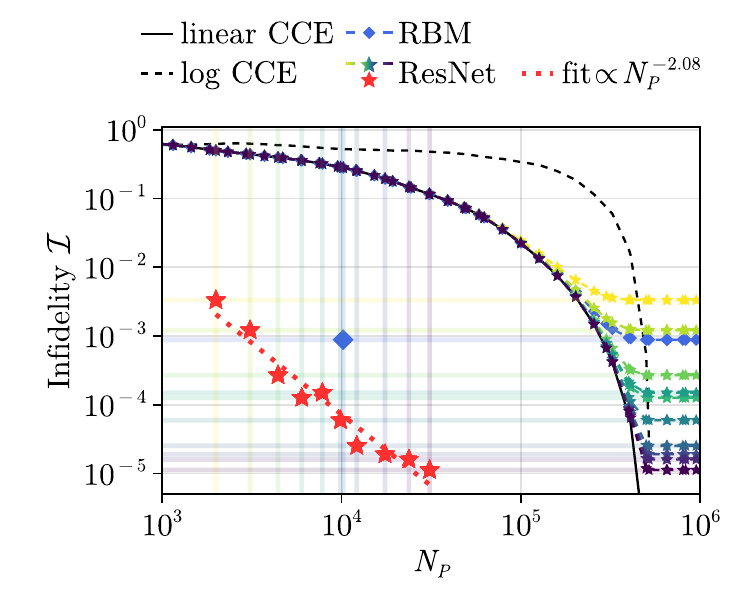}
  \caption{The figure compares the infidelities of the CCE of the exact and variational (ResNet and RBM) states, as a function of the truncation number $N_P$ of the most dominant cumulant contributions, for the critical TFIM at $L=20$ and $tJ=1$. The variational states are ultimately able to reach much lower infidelities (horizontal, colored lines) than the CCE using the same truncation number as the parameter count of the ansatz (vertical, colored lines). The intersection of these lines, marked by red stars, shows the scaling of the infidelity as a function of variational parameters. The dashed line is a power-law fit with exponent $\alpha\simeq -2.08$.}
  \label{fig:representation_scaling}
\end{figure}
For the critical-quench state at $tJ\simeq 1$, Fig.~\ref{fig:representation_scaling} shows a different behavior: both the ResNet and RBM architectures --- with different colors denoting different parameter counts $N_P$, marked by the appropriately colored, vertical lines --- reach significantly infidelities lower than the cumulant ansatz with the same number of parameters --- given by the crossings of the vertical, colored lines with the CCE curves.  

Additionally, we show that a direct Walsh truncation of the linear amplitudes $\psi(\sigma)$ performs better than of the log amplitudes $\log \psi(\sigma)$, but still remains far less efficient than the optimized NQS representations at comparable parameter count. 
This observation contrasts the findings of Ref.~\cite{cortesBasisDependenceNeural2026}, where achieving a given infidelity with NQS and a truncated Walsh decomposition required almost identical parameter numbers. And it is in line with the results of Ref.~\cite{wang2025expressibility} predicting enhanced Walsh complexity of deep NQS. 

The data shown in Fig.~\ref{fig:representation_scaling} also allow us to extract the scaling behavior of the infidelity as a function of $N_P$; the horizontal colored lines show the lowest infidelity reached with the number of parameters denoted by the vertical colored lines. 
The intersection of lines of the same color, marked by red stars, 
illustrates the dependency of the infidelity on the network size $N_P$.
For the ResNet ansatz, we find that the optimized infidelity decreases approximately as a power law $\mathcal{I}\propto N_P^{\alpha}$ with $\alpha\simeq -2.08$ over the tested range of parameter numbers. 
This shows that, at least at representative times beyond the apparent TDVP breakdown, expressivity is not an immediate obstruction.

\section{Continuity in parameter space} \label{sec:continuity}

The supervised optimizations above show that accurate representations of the time-evolved states exist on a finite time grid. 
A stronger requirement to ensure the learned trajectory path can also be found directly by evolving the state using the TDVP equation, is that these representations can be chosen continuously in parameter space. 
We test this by measuring the relative finite-difference velocity
\begin{equation}
    v_\theta(t;\delta t)=\frac{\|\theta(t+\delta t)-\theta(t)\|_2}{\delta t\,\|\theta(t)\|_2}
    \label{eq:param_velocity}
\end{equation}
in the Euclidean metric on the set of parameters found in the previous section. 
While this metric is not invariant under reparameterization of the network, within a fixed architecture it is a direct diagnostic for jumps between disconnected optima; on a continuous path, the velocity should converge to a finite value as the time step $\delta t$ is refined.

\begin{figure}[h!]
  \centering
  \includegraphics[width=\linewidth]{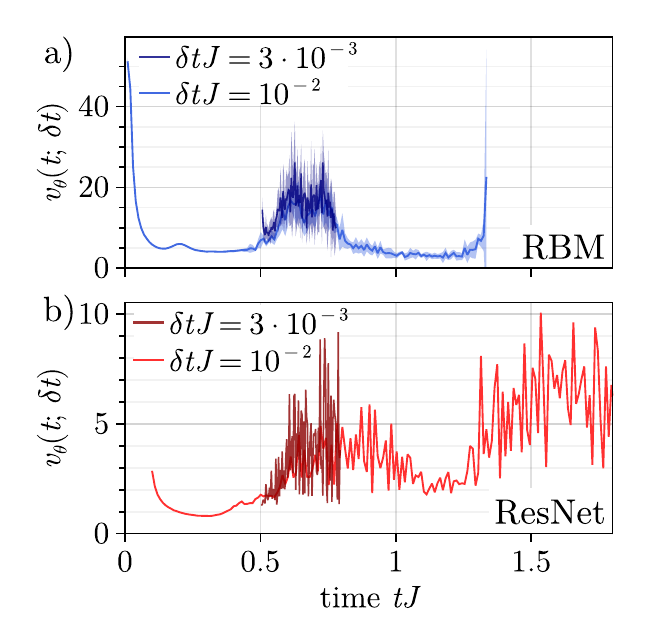}
  \caption{Continuity of the supervised parameter paths for the critical TFIM. The relative finite-difference velocity \(v_\theta(t;\delta t)\) is shown along sequentially optimized a) RBM ($N_P=10240$) and b) ResNet ($N_P=27300$) wave-function paths. Both ans\"atze are compared for \(\delta t=0.01\) and the finer step \(\delta t=0.003\). The ResNet curves show single sequential training paths. For the RBM, solid curves show averages over seven independent random initializations, and the shaded regions denote one standard deviation across the individual RBM runs.}
  \label{fig:param_distance}
\end{figure}
Fig.~\ref{fig:param_distance} shows this diagnostic for the critical TFIM, comparing the found RBM and ResNet paths using two different time steps. 
For both ansätze, the finite-difference velocity remains finite when the time step is refined from $\delta t=0.01$ to $\delta t=0.003$. 
This supports the claim of the existence of a finite time derivative of the parameters and therefore the interpretation that the supervised minima connect continuously as $\delta t \to 0$, rather than forming solutions isolated in time.
The trajectories are nevertheless not uniformly smooth in the chosen parametrization. 
The local spikes and fluctuations in $v_\theta$ can be attributed to details of the supervised training process, such as abort conditions, regularization choices, and optimizer sensitivity.
For the RBM, the curves are averaged over seven random seeds, with shaded regions showing the corresponding spread produced by different runs. 
While the precise height and location of individual spikes vary, the enhanced parameter-space velocity in the time window $0.5 \leq tJ \leq 0.7$ persists across seeds and across both time steps, indicating a reproducible region of large parameter movement, rather than an isolated property of one optimization run.

\section{Optimization} \label{sec:optimization}
So far in this work, we have established the existence of a continuous, high-fidelity path in parameter space $\theta(t)$, which reproduces the time evolution as dictated by the Schrödinger equation with a maximal infidelity of $\mathcal{I}\approx10^{-3}$ for a variety of network architectures, including the ResNet.
In the following, we address the question why the TDVP yields a different trajectory with substantially lower accuracy.

Performing a single TDVP step is an involved problem, comprised of the estimation of the force vector and S-matrix, the regularization required for solving the ill-conditioned linear system, and an adaptive time step integration of the (non-linear) differential equation~\eqref{eq:vmc_param_update}. 
As already mentioned, we employ full summation instead of MC-sampling to ensure a correct, bias-free estimation of all involved quantities.
A comprehensive analysis of the effects of the other hyperparameters is presented in App.~\ref{app:hyperparameter}. 
In summary, while they do impact the quality of the time evolution, the overall signature of a growing infidelity starting at $tJ\approx0.5$, as shown in Fig.~\ref{fig:crit_tfim_failure}~e), persists across all investigated parameter ranges.
The high accuracy solutions found in Sections \ref{sec:expressivity}  and \ref{sec:continuity} cannot be recovered by optimizing the standard hyperparameters. 

\subsection{Time-resolved divergence}

\begin{figure}[t]
    \includegraphics[width=\columnwidth]{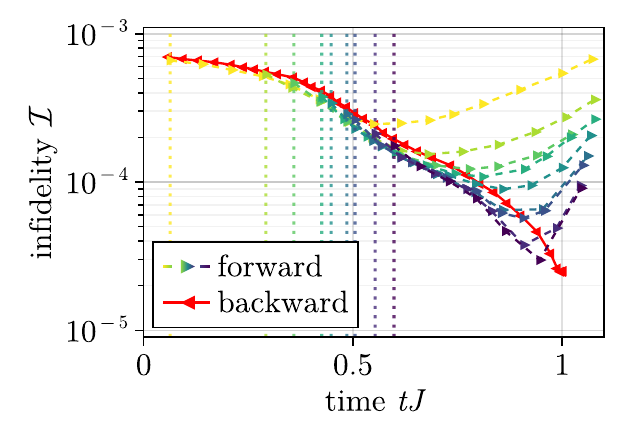}
    \caption{Backward time evolution (red line) starting from the parameters at $tJ=1$ obtained via supervised learning, with subsequent forward evolutions starting at the times marked by vertical lines. The infidelities of the forward simulations increasingly deviate from that of the backward evolution at and after $tJ=0.5$ but agree before this time.} \label{fig:backward_forward}
\end{figure}

To better illuminate the divergence of the TDVP solution, we exploit the accurate representations of time-evolved states at later times generated via supervised learning in Sec.~\ref{sec:expressivity}.
We choose the optimized state $\ket{\psi_{\theta^\star}(t=1)}$, that approximates the exact solution at $tJ=1$ with infidelity $\mathcal I\approx 2\times10^{-5}$, and employ the TDVP to evolve it \emph{backwards} in time.
Thereby, we produce a reference trajectory with high accuracy at a later time point.
The infidelity of this backwards-evolved state with the exact solution displayed in Fig.~\ref{fig:backward_forward}a) furthermore reveals that the backward evolution achieves a higher accuracy compared to the forward evolution of Fig.~\ref{fig:crit_tfim_failure}~e).
While the forward evolution from $tJ=0$ to $tJ=1$ accumulates an infidelity of $\mathcal I\approx10^{-2}$, the infidelity of the backward evolution at $tJ=0$ is $\mathcal I\lesssim10^{-3}$.

Besides establishing a reference trajectory close to the exact solution at late times, this experiment yields additional insights about the difficulty of time-evolving the NQS via TDVP.
The much slower growth of the infidelity during the backward evolution shows that time evolution at later times is not per se hard.
Instead, the forward evolution seems to encounter some kind of bottleneck at the peculiar time $tJ\approx0.5$ and the effect is absent or much less pronounced during backward evolution.
Moreover, this experiment confirms our conclusion from the previous Section \ref{sec:continuity}.
In the backward evolution, TDVP yields a continuous trajectory connecting $\ket{\psi_{\theta(t=0)}}$ and $\ket{\psi_{\theta(t=1)}}$ while achieving high accuracy ($\mathcal I<10^{-3}$) on the whole time interval.

Next, we investigate whether a forward time evolution initialized on the reference trajectory recovers it or whether the forward solution diverges from the reference analogous to Fig.~\ref{fig:crit_tfim_failure}~e).
This is a non-trivial check, since the TDVP equation~\eqref{eq:vmc_param_update}, as opposed to the Schrödinger equation, does not by itself conform to time-reversal symmetry; in particular, it can be broken by the regularization of the S-matrix and potentially through the use of non-holomorphic network architectures.
Further results exploring the time-reversal symmetry of the TDVP equation can be found in App.~\ref{app:time_rev_symm}.

Fig.~\ref{fig:backward_forward} a) shows forward in time simulations starting on the reference trajectory at different points in time, marked by the different colored vertical lines and by the correspondingly colored data points.
Recall that the infidelity shown is with respect to the exact solution.
While in the beginning, all the infidelities of the forward in time evolutions follow the one of the backward in time simulation, they all start to increase compared to the reference trajectory at some point. The trajectories starting at later times diverge later. 
In particular, we show that starting from time $tJ=0$, we are not able to recover the reference trajectory. 
Most noticeably, deviations only start to appear after around $tJ\approx 0.5$ -- before that time, observables of the forward evolutions closely follow ones of the backwards evolution, see also additional data extending to negative $tJ$ in App.~\ref{app:back_forw}. 
This behavior hints at the existence of different stability regimes in the time evolution, i.e. regimes where the physical wave function is affected differently by possible accumulations of errors during the evolution.

\subsection{Sensitivity to perturbations}

\begin{figure}[t]
    \includegraphics[width=\columnwidth]{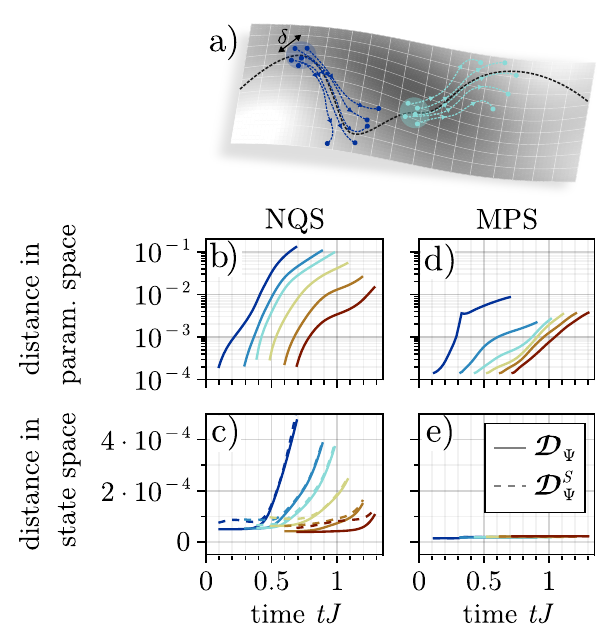}
    \caption{a) Graphical representation of the numerical setup used to test the sensitivity of the TDVP equation by tracking distances between initially nearby trajectories. b)--e) Time evolution of the mean pairwise distance in parameter space (top row) and state space (bottom row). The left column shows results obtained using a ResNet architecture with $N_P=12200$ parameters; the right column shows MPS results with $\chi=12$, optimized via the TDVP equation~\eqref{eq:vmc_param_update}. For both architectures, the distances in parameter space grow quickly. The growth of the NQS distances in state space is delayed, whereas the MPS distances remain nearly constant throughout the time evolution.} \label{fig:sensitivity}
\end{figure}

To investigate a time-dependent sensitivity of the TDVP to perturbations, we initialize the network at different points on the reference trajectory and add random perturbations of strength $\delta$ to the parameters, allowing us to track how these perturbations spread over time, see Fig.~\ref{fig:sensitivity}~a) for a graphical representation.
In order to not drastically change the underlying physical state, the perturbations are taken as random, standard Gaussian vectors $\mathbf r$, normalized to a perturbation strength of $\delta=10^{-4}$.
We investigate $k=5$ random initializations for six different time points, and evolve them for a time of roughly $\Delta t\, J = 0.6$.
Fig.~\ref{fig:sensitivity}~b) shows the relative deviations in parameter-space, defined via
\begin{equation}
    \mathcal{D}_\theta(t) = \frac{2}{k(k-1)} \sum_{i<j}\frac{||\theta_i(t) - \theta_j(t)||}{||\theta_i(t)||},
\end{equation}
with the sum running over different perturbation realizations $i,j$, for the ResNet architecture with $N_P=12200$ parameters.
All trajectories rapidly diverge from each other in parameter space, irrespective of the starting time.
Such sensitivity to perturbations may generally be expected: the TDVP achieves dimensional reduction at the cost of introducing nonlinearity into the equations of motion and, thereby, a potentially chaotic character.
This behavior is in contrast to the observed infidelity of Fig.~\ref{fig:backward_forward}, where a divergence from the reference trajectory only becomes apparent for $tJ\gtrsim0.5$.
However, divergence in parameter space does not necessarily imply divergence of the corresponding wave functions due to possible redundancies in the parameterization.
It is therefore instructive to directly evaluate the mean pairwise trajectory distance of the normalized wave functions, minimized over their relative global phase,
\begin{equation}
    \mathcal{D}_\psi(t)
    = \frac{2}{k(k-1)} \sum_{i<j}\min_{\alpha\in[0,2\pi)}
    \left\|e^{i\alpha}\psi_i(t)-\psi_j(t)\right\|_2\,,
\end{equation}
shown in Fig.~\ref{fig:sensitivity}~c) via solid lines.
While the distances in parameter space grow rapidly immediately after the initialization, the growth of the distances in state space is delayed and depends on the starting time.
In particular for the yellow line, starting close to $tJ=0$, the deviations only start appearing at around $tJ\approx 0.5$, the same time at which we observed the original growth in infidelity~\ref{fig:crit_tfim_failure}~e), while all remaining curves starting at later times begin to increase comparatively earlier.

Furthermore, one can define an alternative measure for the distance in state-space based on the geometry of the variational manifold, by translating the difference in parameter-space $\Delta\theta_{ij}(t) = \theta_i(t) -\theta_j(t)$ to state-space via the QGT $\mathcal{G}$:
\begin{equation}
    \mathcal{D}^{S}_\psi(t) = \frac{2}{k(k-1)}\sum_{i<j}\sqrt{|\Delta \theta_{ij}^T(t) \, \mathcal{G}\,\Delta \theta_{ij}(t)|}\, .
\end{equation}
Note that this only represents a valid approximation to the distance measure if the local structure of the variational manifold at $\theta_i$ and $\theta_j$ is approximately described by the same QGT. 
Conversely, the similarity of the local structure can thus be probed by comparing these two distance measures, as has been done in Fig.~\ref{fig:sensitivity}~c).
We observe the two different state-space distance measures to agree well with each other, confirming that, despite the observed deviations in parameter- and state-space, the corresponding local neighborhoods of the manifold remain similar. 

This supports the existence of different stability regimes in the time evolution: while the variational parameters themselves show a strong sensitivity, with initially nearby trajectories quickly separating from each other, their physical wave functions separate only after a certain simulation time.
Before $tJ\approx0.5$, large parameter-space deviations therefore have only a weak effect on the represented physical states, consistent with perturbations lying predominantly in the approximate null space of the QGT. 
After $tJ\approx0.5$, comparable parameter perturbations produce much larger physical-state deviations, indicating increased overlap with physically relevant directions. 
In the following section, we investigate a possible geometric explanation of this effect.

\subsection{Rotation of the Nullspace}

At this point, it remains unclear whether the observed sensitivity to perturbations is a general consequence of the nonlinear and potentially chaotic nature of the TDVP equation or is tied to the network architecture.
To that end, it is instructive to investigate whether and how this behavior manifests with other variational ansätze, and in the following section, we will be focusing on the well-established MPS. 

While it is not surprising, that the dynamics of the one-dimensional TFIM can be captured using MPS and the single-site TDVP algorithm given a large enough bond dimension, we also find that -- up to numerical discrepancies -- we can reproduce the results and the same level of accuracy when updating the MPS parameters globally via Eq.~\eqref{eq:vmc_param_update}, the same way as has been done for the ResNet in the previous sections. 
In particular, the MPS simulation does not show the same anomalous growth in infidelity at $tJ=0.5$, see App.~\ref{app:mps} for more details.

In order to better understand the origin of the difference between the two variational architectures, we repeat the analysis of the previous section, by initializing the different, randomly perturbed instances of the MPS along its trajectory.
While we still observe rapidly increasing relative deviations in parameter space, deviations in state space remain practically constant throughout the time evolution, independently of the starting time; see Fig.~\ref{fig:sensitivity}~d) and e), respectively.
This result suggests that perturbations, that have been accumulated along nullspace directions of the QGT and which do not affect the physical wavefunction, stay in said nullspace during the time evolution -- which is in stark contrast to the previous observations made for NQS.

In the following, we explain this phenomenology geometrically using two different metrics. 
First, we analyze how the approximate null space of the QGT, and thus the local structure of the physically weakly relevant directions of the variational manifold, evolves in time.
In particular, given the matrix $K_\alpha(t)\in\mathbb{R}^{(p,k)}$, the columns of which contain the basis vectors of the QGT-kernel~\eqref{eq:nullspace} with a threshold of $\alpha$ and at time $t$, we can analyze the rotation of the nullspace between two subsequent TDVP steps via the principal angles $\vartheta_i$, which have been introduced in Ref.~\cite{Bjoerck1973} as
\begin{equation}
    \vartheta_i = \arccos \nu_i\,, \quad \vartheta_i \in [0,\pi/2]\, ,
\end{equation}
where $\nu_i$ denotes the $i$-th singular value of $K(t)^\dag K(t+dt)$. 
However, it has been pointed out in Ref.~\cite{Knyazev2002} that this procedure can produce large round-off errors for small angles, which can be amended through a sine-based correction, see Algorithm 3.1 within said reference, and which we will adopt in this work.
Note that the shape of $K(t)$ does not necessarily remain constant throughout the time evolution, which is why the number of returned principal angles corresponds to the dimension of the smaller null space between $t$ and $t+dt$.
\begin{figure}[t]
    \includegraphics[width=\columnwidth]{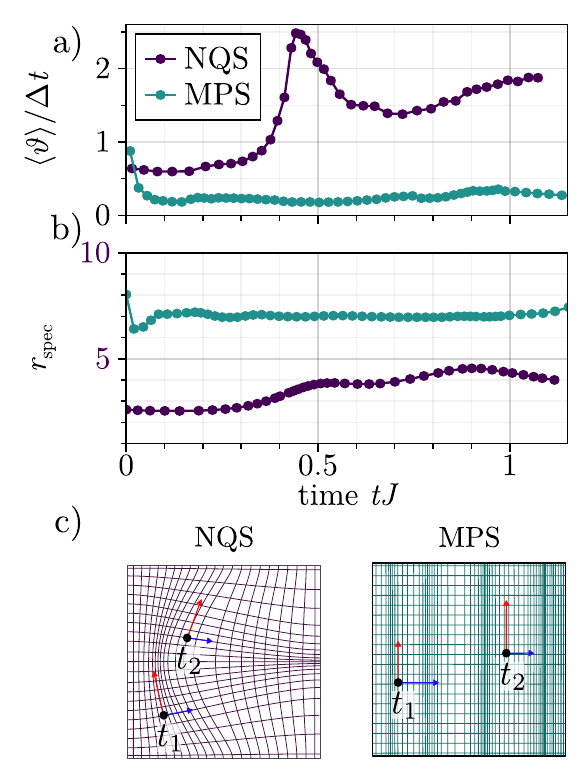}
    \caption{a) Mean principal angle between the null spaces at two subsequent TDVP steps. The angles, and thus the null-space rotation, remain nearly constant and negligible for the MPS, whereas the NQS simulation exhibits strong rotation with a pronounced peak around $tJ\approx0.5$. b) Normalized sum of the absolute eigenvalues, which measures the number of dominant eigendirections on the variational manifold. Both ansätze show the same qualitative features, namely initial growth followed by saturation, but the MPS has roughly five times as many dominant eigendirections as the NQS at each time. c) Schematic manifold-based interpretation: the regular MPS manifold, potentially related to the implicit gauge structure of the architecture, keeps the approximate null-space directions shown in red nearly unchanged. For the NQS, these directions rotate nontrivially between successive times, changing which parameter perturbations weakly or strongly affect the physical state and thereby enabling the amplification of small numerical perturbations.} \label{fig:nullspace_rot}
\end{figure}
Fig.~\ref{fig:nullspace_rot}~a) compares the mean principal angle between the null spaces of variational states separated by roughly $\Delta t\,J\approx0.02$ for an MPS with bond dimension $\chi=12$ and the ResNet architecture used above. We choose the cutoff $\alpha=10^{-8}$, equal to the cutoff used for regularizing the S-matrix.
We observe a stark contrast between the two architectures: the mean principal angle remains nearly constant and negligible throughout the MPS evolution, whereas the NQS angles exhibit pronounced dynamics with a peak at the characteristic time $tJ\approx0.5$.

The dynamics of the principal angles is consistent across different choices of $\Delta t$; see App.~\ref{app:nullspace_rot}. This consistency indicates that the effect is not caused by a finite time step but is related to the intrinsic geometry of the variational manifold.

As a second metric to characterize how the local geometry sampled along the trajectory changes during time evolution, we employ the sum over the normalized, absolute values of the eigenvalues of the QGT 
\begin{equation}
    r_\mathrm{spec}=\sum_i \bigg|\frac{\lambda^{(i)}}{\lambda^{(1)}}\bigg|,
\end{equation}
which is a measure for the number of relevant eigenvectors. 
In Fig.~\ref{fig:nullspace_rot}~b) we observe said quantity to grow in time for both the MPS and NQS simulation, i.e. the number of eigendirections required to correctly capture the dynamics roughly increases with time.
Note however the overall difference in magnitude between the two architectures, the values obtained for MPS by a factor of five larger than ones obtained for NQS.
This can be attributed to the overall very different structures of the eigenspectra, inherent to the two ansätze: while the spectrum for NQS exhibits a smooth decay, the explicit gauge structure of MPS results in a clear separation between active- and nullspace, whereas active directions are utilized very efficiently, resulting in a very flat spectrum within said subspace, see App.~\ref{app:spectra} for more details.

In summary, the NQS exhibits both rapid QGT-null-space rotation and strong physical trajectory amplification in the same time window, whereas the MPS control evolved with the same global TDVP construction exhibits neither. This comparison identifies null-space rotation as the deciding geometric distinction associated with the observed sensitivity. As illustrated schematically in Fig.~\ref{fig:nullspace_rot}~c), the rotation changes which parameter directions weakly or strongly affect the represented state, allowing small perturbations from regularization, finite precision, or integration to produce strongly separated physical trajectories.
By contrast, the MPS exhibits a more rigid manifold structure, which can most likely be attributed to its inherent gauge freedom and has been discussed similarly in Ref.~\cite{wu2026} for another prominent class of tensor-network states, projected entangled-pair states.
While the sum of normalized eigenvalues does not appear to be a decisive metric for understanding the observed difference in evolution quality between MPS and NQS, its growth further characterizes the variation of the local geometry along the trajectory.

\section{Conclusion}\label{sec:conclusion}
This work systematically investigates why applying the TDVP equation to NQS can encounter severe accuracy obstructions for seemingly simple quantum systems already at early times.
We addressed the paradigmatic dynamics in the one-dimensional quantum Ising model upon quenches (close) to the critical point.
In a first step, we have established new empirical upper bounds on the neural network complexity required for accurate representations of the time-evolved states. While confirming an exponential growth of the parameter count with progressing time, the growth rates are much more moderate than what was found in previous work \cite{lin2022scaling}. We attribute this discrepancy to improved optimization within the supervised learning scheme.
Furthermore, we found that the established path in parameter space is continuous, which is a necessary condition for obtaining the same solution by integrating the differential equation prescribed by the infinitesimal TDVP.

Nonetheless, even when circumventing potential challenges related to Monte Carlo estimation and upon careful tuning of the all hyperparameters, we were not able to achieve the accuracy reached in the supervised learning approach by integrating the TDVP equation.
Instead, we have identified a hitherto overlooked challenge related to the TDVP dynamics of non-linear neural network wavefunctions: a sensitivity to small deviations through geometric inflation.
Small perturbations arise naturally from finite precision, numerical integration, and hyperparameter choices such as the regularization cutoff. 
These remain physically irrelevant when aligned with the (approximate) QGT kernel.
However, we observe that the orientation of the QGT kernel changes as the state evolves in time, allowing initially irrelevant perturbations to evolve into strongly separated wave-function trajectories. 
This interpretation is strengthened by comparison with the MPS: the more regular geometry of its manifold, most likely related to its inherent gauge structure, exhibits little null-space rotation and shows no comparable amplification in physical state space.
As a result, the dynamics observed with MPS is very stable and controlled when evolving under the same TDVP.
It remains to be explored whether and how these geometrical properties are related to specific components of NQS, potentially opening the way to engineering nonlinear variational architectures that carry both the expressive power of generic NQS and the stability of tensor-network ansätze. 

The presented diagnostic quantity, given by the mean rotation angle between successive null spaces, can be easily incorporated into existing algorithms and computed cheaply using quantities from the TDVP equation, providing a new tool for identifying potential accuracy obstructions during the time evolution of NQS. 
We expect this mechanism of rotating null spaces not to be restricted to the specific setup and quench considered here, but to apply to a wide range of models and architectures, potentially providing a key ingredient for understanding the difficulties observed in time evolution with NQS.
Nonetheless, the quench to the critical point of the TFIM as investigated in this work constitutes a comprehensive and, due to its simplicity, reproducible testbed for further investigations into the matter of failure modes in the context of time evolution with NQS.

\begin{acknowledgments}
This work was supported via the Helmholtz Initiative and Networking Fund, grant no.~VH-NG-1711, and the DFG Research Unit FOR5919 ``Machine Learning for Complex Quantum States", grant no.~550495627.
The authors gratefully acknowledge the Gauss Centre for Supercomputing e.V. (www.gauss-centre.eu) for funding this project by providing computing time through the John von Neumann Institute for Computing (NIC) on the GCS Supercomputer JUWELS at Jülich Supercomputing Centre (JSC)
and the scientific support and HPC resources provided by the Erlangen National High Performance Computing Center (NHR@FAU) of the Friedrich-Alexander-Universität Erlangen-Nürnberg (FAU) under the NHR project b245da / JA-22555. NHR funding is provided by federal and Bavarian state authorities. NHR@FAU hardware is partially funded by the German Research Foundation (DFG) – 440719683. 
Numerical simulations were implemented based on the jVMC codebase \cite{Schmitt_2022}.
Data and code are available at \cite{krinitsin_2026_data}.
\\
\end{acknowledgments}

\newpage
\appendix


\section{Network Architectures} \label{app:networks}

We use three NQS architectures: a complex restricted Boltzmann machine (RBM)~\cite{carleo,melko2019}, an autoregressive recurrent neural network (RNN) inspired by Ref.~\cite{hibat2020recurrent}, and a convolutional residual network (ResNet) following Ref.~\cite{chen2024empowering}. Table~\ref{tab:network_architectures} summarizes the depth and width of each network and, in particular, how its size was increased in the expressivity analysis.

\begin{table*}[t]
\centering
\small
\setlength{\tabcolsep}{4pt}
\renewcommand{\arraystretch}{1.2}
\begin{tabular}{@{}p{0.15\textwidth}p{0.25\textwidth}p{0.28\textwidth}p{0.25\textwidth}@{}}
\hline\hline
 & \textbf{RBM} & \textbf{RNN} & \textbf{ResNet} \tabularnewline
\hline
Structure
& $\sigma\in\{-1,1\}^{L}$, $W\in\mathbb{C}^{N_h\times L}$,
  \newline $(\sigma,W)\mapsto\log\psi_\theta(\sigma)\in\mathbb{C}$.
& At site $i$, $\sigma_i\in\{0,1\}$ is represented by $e_{\sigma_i}\in\mathbb{R}^{d}$ and mapped to $x_i=E e_{\sigma_i}+e_0\in\mathbb{R}^{d_h}$, with $E\in\mathbb{R}^{d_h\times d}$ and $e_0\in\mathbb{R}^{d_h}$.
  \newline For the gates $q\in\{f,i,o,g\}$, $W_q,U_q\in\mathbb{R}^{d_h\times d_h}$ and $b_q\in\mathbb{R}^{d_h}$,
  \newline $(x_i,y_{i-1},m_{i-1})\mapsto(y_i,m_i)$.
  \newline The two logits are $\ell_i=Ay_i+a\in\mathbb{R}^{d}$, with $A\in\mathbb{R}^{d\times d_h}$ and $a\in\mathbb{R}^{d}$, and $\log p_i=\log\operatorname{softmax}(\ell_i)$.
  \newline For $d_h\leq32$, $\phi:(y_L,w_\phi,b_\phi)\mapsto\mathbb{R}$, with $w_\phi\in\mathbb{R}^{d_h}$ and $b_\phi\in\mathbb{R}$. For $d_h>32$, $\phi$ additionally contains $V_\phi\in\mathbb{R}^{d_h\times d_h}$ and $v_\phi\in\mathbb{R}^{d_h}$.
  \newline $\displaystyle \log\psi_\theta(\sigma)=\frac12\sum_{i=1}^{L}\log p_{i,\sigma_i}+i\phi$.
& $X^{(0)}_{i,1}=\sigma_i$, $K^{(1)}\in\mathbb{R}^{F\times1\times c}$, and $K^{(\ell)}\in\mathbb{R}^{F\times c\times c}$ for $2\leq\ell\leq D$.
  \newline Each residual block $b=1,2$ produces
  \newline $X^{(2b)}=X^{(2b-2)}+\mathcal{F}_b\!\left(X^{(2b-2)};K^{(2b-1)},K^{(2b)}\right)\in\mathbb{R}^{L\times c}$,
  \newline where $\mathcal{F}_b$ is the two-convolution nonlinear branch and $X^{(0)}$ is broadcast over the $c$ channels in the first residual addition. 
  \newline The final fields form
  \newline $z_{i,a}=X^{(D)}_{i,a}+iX^{(D)}_{i,a+c/2}$, $1\leq a\leq c/2$,
  \newline $\displaystyle \log\psi_\theta(\sigma)=\log\!\sum_{i,a}e^{z_{i,a}}-\log(cL)$. \tabularnewline
\hline

Fixed dimensions
& $L=20$; one hidden layer; no bias parameters
& $L=20$, $d=2$; one recurrent layer, with the same transformation applied at every site
& $L=20$, $D=4$ convolutional layers in two residual blocks, $F=10$; unit stride, zero padding, and no bias parameters \tabularnewline

Scaled dimension
& $N_h$, the number of complex hidden units
& $d_h$, the dimension of the embedding and recurrent state vectors
& Even-integer valued $c$, the number of real channels, increased simultaneously in all $D$ layers \tabularnewline

Parameter scaling
& $N_P=2LN_h=40N_h$
& NIIM: $N_P=8d_h^2+10d_h+3$.
  \newline TFIM: $N_P=8d_h^2+10d_h+4$ for $d_h\leq32$, and $N_P=9d_h^2+11d_h+4$ for $d_h>32$.
& $N_P=Fc+(D-1)Fc^2=10c+30c^2$ \tabularnewline
\hline
Reference
& Refs.~\cite{carleo,melko2019}
& Ref.~\cite{hibat2020recurrent}
& Ref.~\cite{chen2024empowering} \tabularnewline
\hline\hline
\end{tabular}
\caption{Network dimensions and size scaling used in the supervised expressivity study. Here, $y_i$ and $m_i$ denote the output and memory vectors of the RNN, respectively. All size scans vary one width parameter while keeping the remaining dimensions fixed.}
\label{tab:network_architectures}
\end{table*}

The NIIM calculations use the linear RNN phase map throughout. For the wider RNNs required by the TFIM, the additional nonlinear phase layer is included when $d_h>32$. Throughout Sec.~\ref{sec:optimization}, the ResNet depth remains fixed at two residual blocks and we use $c=20$, corresponding to $N_P=12200$ real parameters.

\section{Supervised Learning Approach} \label{app:sup_learning}

\subsection{Infidelity as Loss Function}

The supervised training approach requires a measure of distinguishability between the target wave function $\ket{\chi_t}$ and the NQS $\ket{\psi_{\theta_t}}$, such that this measure $\text{dist}\big(\ket{\chi_t},\ket{\psi_{\theta_t}}\big) = 0$ when $\ket{\psi_{\theta_t}} \propto \ket{\chi_t}$. For this purpose, we adopt the infidelity, which is defined as
\begin{equation}
\mathcal{I} =
1-\frac{|\braket{\psi_\theta|\chi}|^2}
{\braket{\psi_\theta|\psi_\theta}\braket{\chi|\chi}}\;,
\end{equation}
since it was recently shown in Ref.~\cite{sinibaldi2023unbiasing} to be advantageous over other measures in combination with VQMC. In that spirit, the expression \eqref{eq:infidelity} can be rewritten as an expectation value
\begin{equation}
\label{eq:sample_inf}
\mathcal{I} =
\big\llangle\mathcal{I}_{\rm loc}(\sigma,\eta)\big\rrangle_{\Pi}
\end{equation}
where we have introduced the local infidelity estimator
\begin{equation}
\mathcal{I}_{\rm loc}(\sigma,\eta) = 1- \frac{\braket{\sigma|\chi}\braket{\eta|\psi_\theta}}{\braket{\sigma|\psi_\theta}\braket{\eta|\chi}} \;.
\end{equation}
and the product probability distribution
\begin{align}
\Pi(\sigma,\eta) = \frac{\big|\psi_\theta(\sigma)\big|^2\big|\chi(\eta)\big|^2}{\braket{\chi|\chi}\braket{\psi_\theta|\psi_\theta}}\;,
\end{align}
which consists of the Born distributions of the NQS and the target wave function respectively 
\begin{equation}
\pi_\psi(\sigma) = \frac{\big|\psi_\theta(\sigma)\big|^2}{\braket{\psi_\theta|\psi_\theta}} \quad
\pi_\chi(\eta) = \frac{\big|\chi(\eta)\big|^2}{\braket{\chi|\chi}} \;.
\end{equation}

\subsection{Estimate the infidelity}

It was shown by Sinibaldi et al. in Ref.~\cite{sinibaldi2023unbiasing} that the signal to noise ratio (SNR) of $\mathcal{I}_{\rm loc}$ vanishes as 
\begin{equation}
\text{SNR}[\mathcal{I}_{\rm loc}]
=\sqrt{N_{MC}}\,
\frac{\big|\big\llangle\mathcal{I}_{\rm loc}\big\rrangle_{\Pi}\big|}
{\sqrt{\text{var}(\mathcal{I}_{\rm loc})}}
\propto \sqrt{\mathcal{I}} \;.
\end{equation}
However, Sinibaldi et al. have further shown that this can be counteracted by applying the \textit{control variates} (CV) principle. Defining the residual $R=1-\mathcal{I}_{\rm loc}$, the estimator is
\begin{equation}
\mathcal{I}^{CV}_{\rm loc}(\eta,\sigma) =
\mathcal{I}_{\rm loc}(\eta,\sigma)
+ c \Big\{|R|^2-\big\llangle|R|^2\big\rrangle_{\Pi}\Big\} \;,
\end{equation}
which can be conveniently simplified because $\big\llangle|R|^2\big\rrangle_{\Pi}=1$. 
With the optimal CV parameter
\begin{equation}
c = -\frac{\operatorname{cov}\big[\operatorname{Re}R,|R|^2\big]}
{\operatorname{var}\big[|R|^2\big]}
\end{equation}
it can be shown that the SNR remains stable $\text{SNR}[\mathcal{I}_{\rm loc}] \propto \mathcal{O}(1)$ as the infidelity vanishes.
To avoid having to recalculate the CV parameter, it is convenient to set it to its limit value $\mathcal{I} \to 0: c \to -1/2$ at all times. 
The control-variate term has zero expectation, so fixing $c=-1/2$ does not bias the infidelity estimate. Individual finite-sample estimates can nevertheless fluctuate outside the interval $[0,1]$ when $\mathcal{I}\approx1$.

\subsection{Infidelity gradient}

With a distinguishability measure at hand, the optimal representation of $\ket{\chi_t}$ by $\ket{\psi_\theta}$ can be found by minimizing Eq.~\eqref{eq:sample_inf}.
The infidelity is minimized using \textit{gradient descent} (GD), which requires the estimation of the gradient through MC sampling for the optimisation procedure. 
With \(O_k(\sigma)=\partial_{\theta_k}\log\psi_\theta(\sigma)\), the gradient of \ceqn{eq:sample_inf} can be obtained by applying the product rule, which gives a contribution from the derivative of $\Pi(\sigma,\eta)$ and a contribution from $\mathcal{I}_{\rm loc}(\sigma,\eta)$
\begin{widetext}
\begin{align}
\label{eq:naiv_grad}
\partial_{\theta_k}\mathcal{I}
&=\Re\Big\{
2\big\llangle \Re\, O_k(\sigma)\mathcal{I}_{\rm loc}(\sigma,\eta)\big\rrangle_\Pi
-2\big\llangle \Re\, O_k(\sigma)\big\rrangle_\Pi
\big\llangle\mathcal{I}_{\rm loc}(\sigma,\eta)\big\rrangle_\Pi
\nonumber\\
&\quad
+\Big\llangle
\big(O_k(\sigma)-O_k(\eta)\big)
\big(1-\mathcal{I}_{\rm loc}(\sigma,\eta)\big)
\Big\rrangle_\Pi
\Big\}\;,
\end{align}
\end{widetext}
Similarly to the bare infidelity, the gradient has a SNR that vanishes as the infidelity vanishes.

Control variates stabilize the sampled infidelity itself, but differentiating the control-variate estimator does not automatically yield the most stable gradient estimator. 
A systematic comparison of related stochastic infidelity-gradient estimators in Ref.~\cite{gravinaNeuralProjectedQuantum2025a} identifies the covariance structure as the key ingredient for favorable SNR and compatibility with minSR. 
In our notation, the same gradient can be written as
\begin{align}
\label{eq:inf_gradient}
\partial_{\theta_k}\mathcal{I}
&=2\Re\Big[
\big\llangle O_k(\sigma)\mathcal{I}_{\rm loc}(\sigma,\eta)\big\rrangle_\Pi
\nonumber\\
&\quad
-\big\llangle O_k(\sigma)\big\rrangle_\Pi
\big\llangle\mathcal{I}_{\rm loc}(\sigma,\eta)\big\rrangle_\Pi
\Big]\;.
\end{align}
Comparing the SNR and the optimization results between both formulations of the gradient, \ceqn{eq:inf_gradient} is found to have a vastly larger SNR and to produce lower infidelities than \ceqn{eq:naiv_grad}.

For completeness, we derive the covariance form in Eq.~\eqref{eq:inf_gradient}. We consider
\begin{equation*}
\Pi(\sigma,\eta)\propto
|\psi(\sigma)|^2|\chi(\eta)|^2,
\qquad
\sigma\sim\psi,\;\eta\sim\chi .
\end{equation*}
We start from
\begin{align*}
|\braket{\psi|\chi}|^2
&=
\sum_{\sigma,\eta}
\psi^*(\sigma)\chi(\sigma)
\chi^*(\eta)\psi(\eta)
\\
&=
\sum_{\sigma,\eta}
|\psi(\sigma)|^2|\chi(\eta)|^2
\frac{\chi(\sigma)}{\psi(\sigma)}
\frac{\psi(\eta)}{\chi(\eta)} .
\end{align*}
Equivalently,
\begin{align*}
|\braket{\psi|\chi}|^2
&=
\sum_{\sigma,\eta}
|\psi(\eta)|^2|\chi(\sigma)|^2
\\
&\quad\times
\frac{\psi^*(\sigma)}{\chi^*(\sigma)}
\frac{\chi^*(\eta)}{\psi^*(\eta)} .
\end{align*}
Together with the derivative of the norm,
\begin{align*}
\partial_{\theta_k}
&\left[
\frac{\braket{\psi|\sigma}\braket{\eta|\psi}}
{\braket{\psi|\psi}}
\right]
\\
&=
\frac{\braket{\partial_{\theta_k}\psi|\sigma}
\braket{\eta|\psi}
+\braket{\psi|\sigma}
\braket{\eta|\partial_{\theta_k}\psi}}
{\braket{\psi|\psi}}
\\
&\quad
-2\,\mathbb{E}_\psi[\Re\,O_k]\,
\frac{\braket{\psi|\sigma}\braket{\eta|\psi}}
{\braket{\psi|\psi}}
\\
&=
\frac{\braket{\psi|\sigma}\braket{\eta|\psi}}
{\braket{\psi|\psi}}
\Big[
O_k^*(\sigma)+O_k(\eta)
-2\,\mathbb{E}_\psi[\Re\,O_k]
\Big],
\end{align*}
where \(O_k(\sigma)=\partial_{\theta_k}\log\psi_\theta(\sigma)\).
Using the conjugate form of the local estimator gives
\begin{align*}
\partial_{\theta_k}
\big\llangle\mathcal{I}_{\rm loc}\big\rrangle_\Pi
&=
\sum_{\sigma,\eta}\Pi(\sigma,\eta)
\Big[
\mathcal{I}_{\rm loc}(\sigma,\eta)O_k(\sigma)
\\
&\quad
+\mathcal{I}_{\rm loc}^*(\sigma,\eta)O_k^*(\sigma)
\\
&\quad
-2\,\mathbb{E}_\psi[\Re\,O_k]\,
\mathcal{I}_{\rm loc}(\sigma,\eta)
\Big].
\end{align*}
Collecting terms, we obtain the covariance structure
\begin{align*}
\partial_{\theta_k}\mathcal I
&=
2\,\Re\Big[
\big\llangle
O_k(\sigma)\mathcal I_{\rm loc}(\sigma,\eta)
\big\rrangle_\Pi
\\
&\quad
-\big\llangle O_k(\sigma)\big\rrangle_\Pi
\big\llangle
\mathcal I_{\rm loc}(\sigma,\eta)
\big\rrangle_\Pi
\Big].
\end{align*}

\subsection{Infidelity optimization}

The infidelity is optimized with a natural-gradient step built from the covariance estimator in Eq.~\eqref{eq:inf_gradient}. 
In practice we use its sample-space form, often referred to as minSR, which replaces the inversion of the parameter-space quantum geometric tensor by the inversion of the neural tangent kernel in the Monte Carlo sample space \cite{chen2024empowering}. 
This is appropriate for the present gradient estimator, since its covariance structure reduces stochastic fluctuations and makes the force compatible with the minSR linear system.

The diagonal shift entering this linear system is not kept fixed. 
Following the Levenberg--Marquardt logic recently used for stochastic infidelity minimization in projected quantum dynamics \cite{gravinaNeuralProjectedQuantum2025a}, each proposed step is tested by comparing the measured change in the sampled infidelity with the quadratic prediction of the local model. 
The ratio is used to update the shift stored for the subsequent optimization iteration. 
The gradient clipping used here is an additional stabilization beyond the adaptive damping of Ref.~\cite{gravinaNeuralProjectedQuantum2025a}. 
After each minSR solve, we compute the norm of the proposed parameter-space update and compare it with a running mean of the norms of previously accepted updates. 
Once this running scale has been initialized, an anomalously large update, here larger than ten times the running mean, or an almost vanishing update is rejected on the first trial and recomputed before acceptance. 
Accepted updates are then rescaled by at most the ratio between the running mean and their norm, leaving smaller updates unchanged. 
This simple history-dependent clipping removes rare Monte Carlo spikes without introducing an additional fixed trust-region scale, and was essential for stable optimization over long training trajectories.
%

For implementation we define
\begin{equation}
\bar O_{\sigma,k}=O_k(\sigma)-\big\llangle O_k\big\rrangle_{|\psi|^2},
\end{equation}
and
\begin{equation}
\bar{\mathcal I}_\sigma=\big\llangle\mathcal I_{\rm loc}\big\rrangle_{\eta \sim|\chi|^2}
- \big\llangle\mathcal I_{\rm loc}\big\rrangle_{(\sigma,\eta)\sim\Pi}.
\end{equation}
For real variational parameters, these complex quantities are represented by
the real matrices
\begin{equation}
X=\frac{1}{\sqrt{N_{\rm samples}}}
\begin{pmatrix}
\operatorname{Re}\bar O\\
\operatorname{Im}\bar O
\end{pmatrix},
\qquad
e=\frac{2}{\sqrt{N_{\rm samples}}}
\begin{pmatrix}
\operatorname{Re}\bar{\mathcal I}\\
\operatorname{Im}\bar {\mathcal I}
\end{pmatrix}.
\end{equation}
Thus, \(X\in\mathbb{R}^{2N_{\rm samples}\times N_{\rm param}}\),
\(e\in\mathbb{R}^{2N_{\rm samples}}\), and
\(T=XX^{\mathsf T}\) is the sample-space kernel. The minSR direction is
\begin{equation}
\delta_\lambda
=
X^{\mathsf T}
\left(T+\lambda\mathds{1}\right)^{-1}
e,
\qquad
\theta^\prime=\theta-\eta\,\delta_\lambda .
\end{equation}

The Levenberg--Marquardt ratio is evaluated from the sampled loss and the same local quadratic model,
\begin{equation}
\rho = \frac{\mathcal{I}(\theta^\prime)-\mathcal{I}(\theta)}{\nabla\mathcal{I}^{\mathsf T} \Delta\theta + \frac{1}{2}\Delta\theta^{\mathsf T} X^{\mathsf T} X\Delta\theta}, \qquad \Delta\theta=\theta^\prime-\theta .
\end{equation}
For the subsequent iteration, we update the stored shift according to
\begin{equation}
\lambda\leftarrow
\begin{cases}
\max(0.1\lambda,10^{-12}), & \rho>\rho_{\max},\\
\lambda, & \rho_{\min}<\rho\leq\rho_{\max},\\
\min(10\lambda,10^{-4}), & \rho\leq\rho_{\min}.
\end{cases}
\end{equation}
In the last case, the current direction is accepted subject to the history-based clipping below rather than being recomputed with the updated shift.
The clipping scale is
\begin{equation}
\bar g_n=\frac{1}{M}\sum_{m=n-M}^{n-1}\|\delta_m\|,
\end{equation}
where only accepted updates enter the running history. On the first trial after initialization, the step is rejected if
\begin{equation}
\|\delta_\lambda\|>a\,\bar g_n \quad \text{or} \quad \|\delta_\lambda\|<g_{\min}.
\end{equation}
Otherwise the accepted direction is clipped according to
\begin{equation}
\delta_\lambda \leftarrow \delta_\lambda\min\left(1,\frac{\bar g_n}{\|\delta_\lambda\|}\right).
\end{equation}
The numerical values used here were \(M=20\), \(a=10\), \(g_{\min}=10^{-10}\), \(\rho_{\min}=0.25\), \(\rho_{\max}=0.75\), and \(\lambda_0=10^{-3}\). In the poor-agreement branch, the updated shift was capped at \(10^{-4}\).

\section{Investigation of TDVP Hyperparameters}\label{app:hyperparameter}
As a starting point, we investigate the dependence of the time evolution on the regularization parameter $\epsilon$, see~\eqref{eq:lambda_inv}, after initializing the state with the parameters found through infidelity minimization at $tJ=0.4$, ensuring the state to lie on one of the potentially optimal paths.
\begin{figure}[h!]
    \includegraphics[width=0.9\columnwidth]{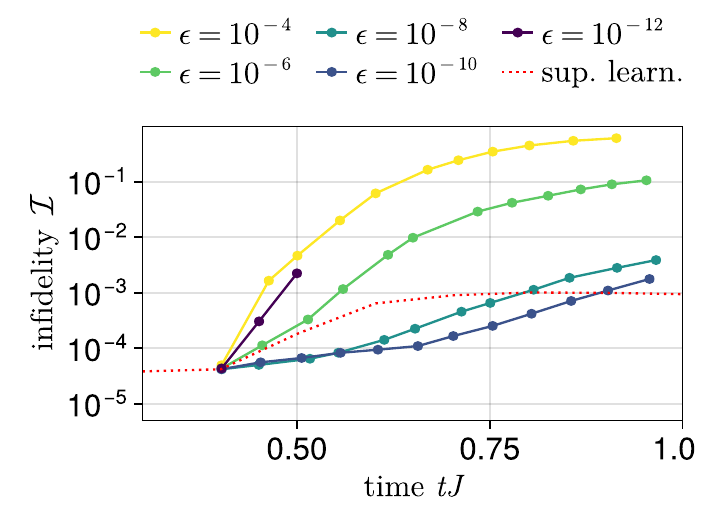}
    \caption{This figure shows the infidelity growth for different regularization parameters $\epsilon$, starting with the parameters obtained from supervised learning of the exact state at time $tJ=0.4$. The red dotted line denotes the infidelity from said learned states. We observe a cutoff parameter in the range $10^{-10} \leq \epsilon \leq 10^{-8}$ to provide the best results.} \label{fig:cutoff}
\end{figure}
By tracking the infidelity to the exact reference result, we find that there is an optimal and robust region for the cutoff parameter $10^{-10} \leq \epsilon \leq 10^{-8}$, see Fig.~\ref{fig:cutoff}.
As such, we set $\epsilon=10^{-8}$ throughout all simulations.

\section{Time-reversal symmetry of the TDVP equation}\label{app:time_rev_symm}

In this section we discuss analytic considerations and numerical results regarding the time-reversal symmetry of the TDVP equation. 
While the Schrödinger equation naturally respects time reversal symmetry for Hermitian Hamiltonians, this property is not straightforwardly inherited by variational methods such as TDVP. 
In particular, the variational manifold $\mathcal{M}$ has to be closed under the time-reversal transformation $\mathcal{T}\mathcal{M}=\mathcal{M}$.

Time-reversal symmetry implies that the time-reversed trajectory $\bar{\theta}(t)\equiv\mathcal{T}\theta=\theta^*(-t)$ satisfies the same TDVP equation of motion~\eqref{eq:vmc_param_update}. 
Considering the explicit form of the QGT~\eqref{eq:Q} and force vector~\eqref{eq:F}, one readily observes that this is the case if the network satisfies $\psi_{\theta^*}(\sigma)=\psi^*_\theta(\sigma)$, which is true for holomorphic neural networks. 
Note, however, that this does not imply that non-holomorphic networks automatically break time-reversal symmetry, since the above condition can also hold in more general cases.
An additional complication is given by the regularization, which depends on the local structure of the manifold, and as such can differ when performing a forward or backward in time step between two points, breaking time reversal symmetry. 
To check the influence of the regularization, we can solve the equation of motion in an alternative fashion by assuming a complex and holomorphic network parametrization: instead of solving Eq.~\eqref{eq:vmc_param_update}, we can utilize the fact that the QGT and force vector take on the form $\mathcal{G}=\bar{O}^\dag \bar{O}$ and $\mathcal{F}=\bar{O}^\dag \bar{E}_\mathrm{loc}$, respectively, with $E_\mathrm{loc}$ the local energy and centered quantities denoted by bars.
The normal equation is equivalently the optimality condition for the least-squares problem
\begin{align}
    \bar{O}^*_{k,\sigma} \bar{O}_{\sigma,k'}\, \dot\theta_{k'}
    &= \bar{O}^*_{k,\sigma} \bar{E}_{\mathrm{loc},\sigma} \nonumber \\
    \dot\theta
    &= \operatorname*{arg\,min}_{v}\|\bar{O} v-\bar E_\mathrm{loc}\|_2^2.
    \label{eq:jac_param_update}
\end{align}
We solve the latter problem through a singular-value decomposition of the centered Jacobian $\bar{O}_{\sigma,k'}$. 
Its singular values are exactly the square roots of the nonzero eigenvalues of the QGT $\mathcal{G}=\bar{O}^\dag \bar{O}$. 
They therefore cover a smaller dynamic range than the QGT eigenvalues. In the numerical test below, we solve the least-squares problem by SVD without applying the S-matrix soft cutoff.
In the following, we will refer to this approach as Jacobian inversion.

\begin{figure}[h!]
    \includegraphics[width=\columnwidth]{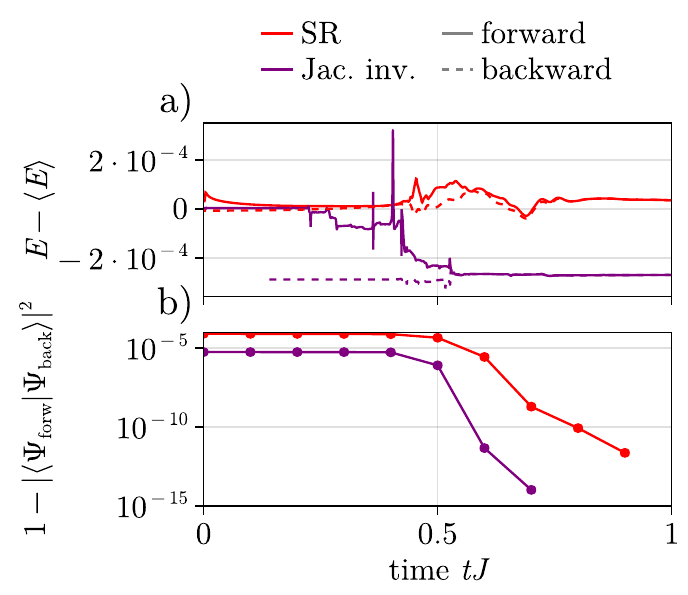}
    \caption{Comparison of a forward and subsequent backward time evolution using the regularized S-matrix equation~\eqref{eq:vmc_param_update} and the unregularized least-squares formulation~\eqref{eq:jac_param_update}. a) Energies along the different trajectories. b) Infidelity between the forward and backward trajectories. The regularized S-matrix equation produces faster infidelity growth than the least-squares formulation used without the S-matrix soft cutoff.} \label{fig:woS_backward_forward}
\end{figure}

In Fig.~\ref{fig:woS_backward_forward}, we investigate time reversibility and the impact of regularization numerically for a quench to the critical point of the TFIM on a one-dimensional chain of size $L=16$, using a holomorphic RBM as our variational ansatz.
Fig.~\ref{fig:woS_backward_forward}a) compares the evolution of an observable, the energy, between the different trajectories, showing marginal discrepancies between the forward and backward evolution.

A more sensitive metric is the infidelity between the forward and backward evolutions, shown in Fig.~\ref{fig:woS_backward_forward}~b). Solving the unregularized equation leads to infidelities smaller by several orders of magnitude than the regularized version, confirming the impact of regularization on the time-reversal symmetry of the TDVP equation. Nevertheless, both infidelities grow significantly over the observed time period.
This behavior is consistent with the sensitivity discussed in the main text: finite-precision and rounding perturbations can be amplified as the approximate QGT null space rotates, spoiling time-reversal symmetry even without regularization artifacts.

\section{Further data for the back-/and forward evolution}\label{app:back_forw}

\begin{figure}[h!]
    \includegraphics[width=\columnwidth]{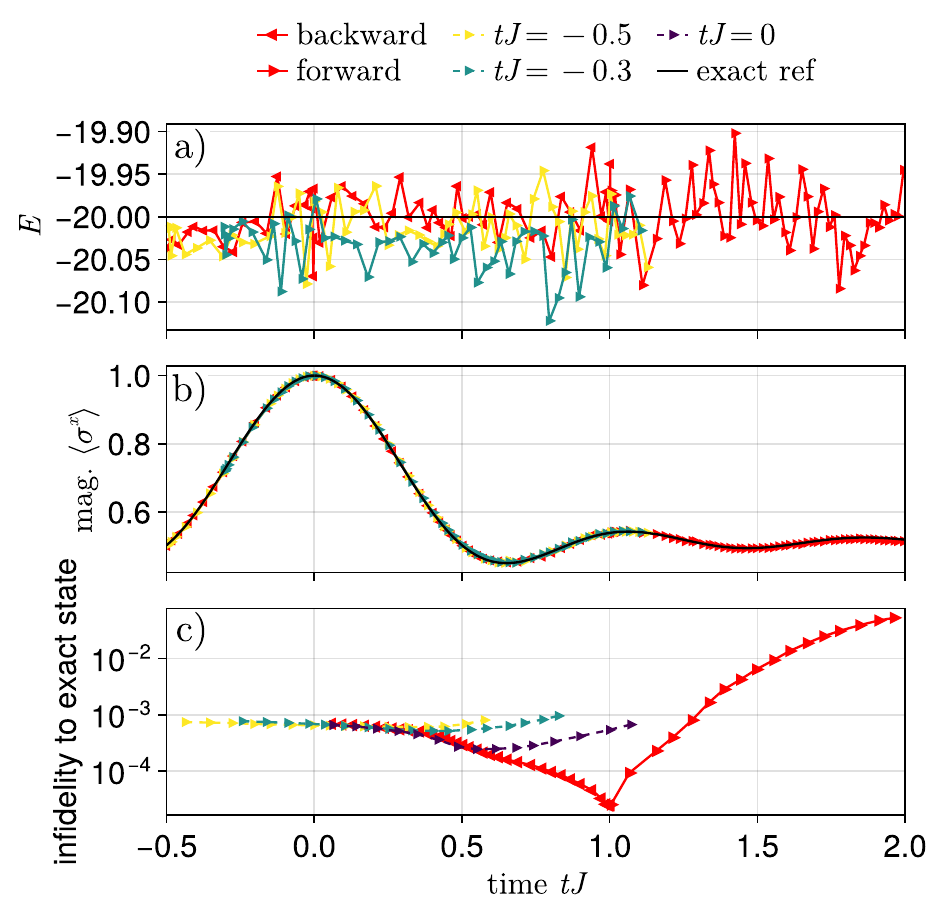}
    \caption{Additional forward and backward trajectories extending the regime in both negative and positive time directions. a) Time evolution of the energy. b) Time evolution of the magnetization. c) Time evolution of the infidelity with respect to the exact state.} \label{fig:backward_forward_app}
\end{figure}

Fig.~\ref{fig:backward_forward_app} shows additional data regarding the analysis of the back- and subsequent forward in time evolutions, starting from the well-converged parameters at $tJ=1$ obtained through infidelity minimization. 
We extend the backward in time evolution to negative times, up to $tJ=-0.5$, allowing us to better investigate the different stability regimes.
Starting from the parameters at $tJ=-0.5$ and $tJ=-0.3$ on the backward-evolved curve, we observe that both the energy and magnetization follow the reference curve. The infidelity with respect to the exact state agrees between the forward and backward evolutions up to the characteristic time $tJ\approx0.4$, where the curves start to diverge, providing further support for the existence of different stability regimes.

Furthermore, starting from the well converged parameters at $tJ=1$, we also provide data for the forward in time evolution. 
Interestingly, the infidelity exhibits an even faster growth compared to the backward evolution, which, comparing to the expressivity analysis in Fig.~\ref{fig:expressivity_summary}, may be related to the saturation of the expressive power of the ResNet at $N_P=12200$ parameters in time.

\section{Time-evolution using MPS} \label{app:mps}
Even without explicitly utilizing the gauge structure to reformulate the TDVP equation in terms of a sequence of local updates, MPS can reliably represent the state during the time evolution after a quench to the critical TFIM when updated globally using Eq.~\eqref{eq:vmc_param_update}.
\begin{figure}[h!]
    \includegraphics[width=\columnwidth]{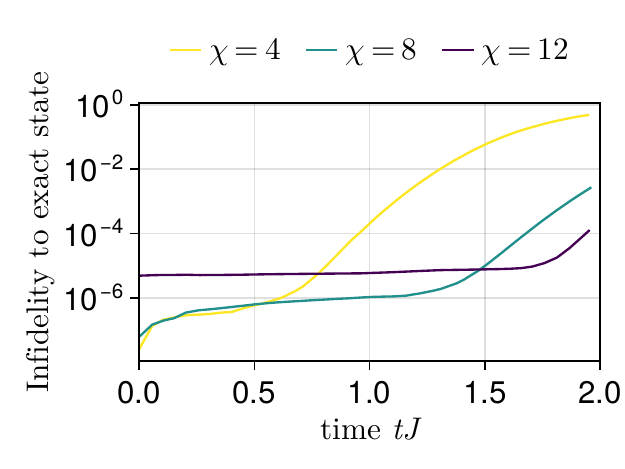}
    \caption{Time evolution of the infidelity to the exact state using MPS and the global TDVP parameter update~\eqref{eq:vmc_param_update}. While an MPS with a bond dimension of $\chi=4$ fails to correctly capture the state due to a limited expressivity, increasing the bond dimension systematically improves its capacity and accuracy.} \label{fig:mps_app}
\end{figure}
Fig.~\ref{fig:mps_app} shows the infidelity to the exact state for different bond dimensions: the limitations of the expressivity at a finite bond dimension can be clearly seen, while increasing the bond dimension allows for a systematic way to improve the accuracy.

\section{Comparing the eigenspectrum of the S-matrix between MPS and NQS} \label{app:spectra}

\begin{figure}[h!]
    \includegraphics[width=\columnwidth]{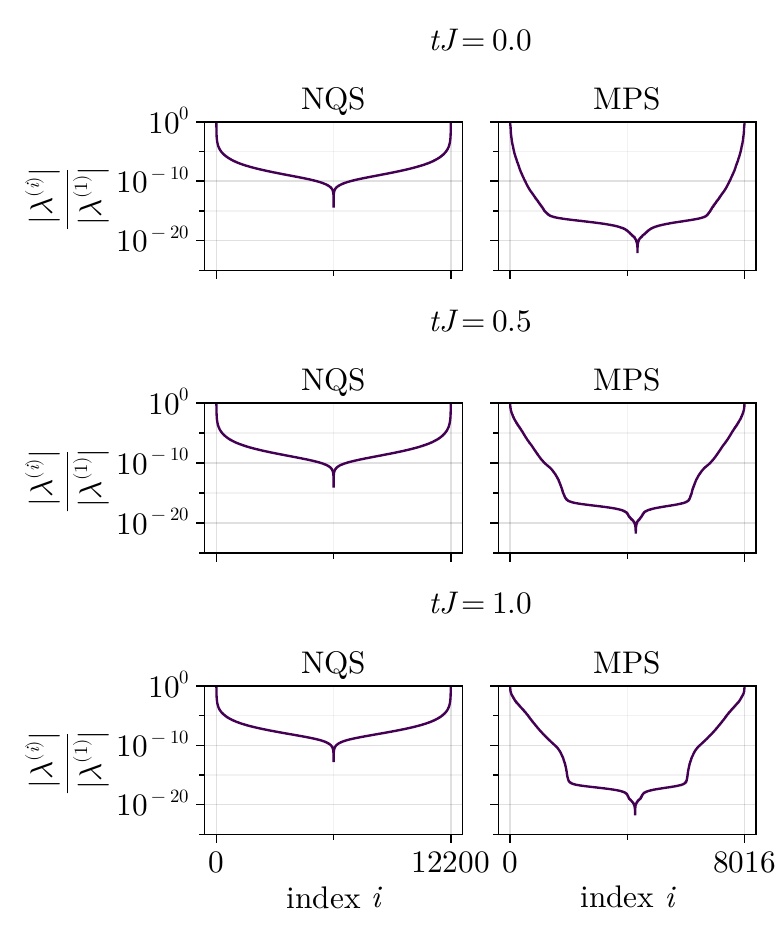}
    \caption{Normalized spectra of the Hermitian S-matrix, $S=i\operatorname{Im}\mathcal{G}$, for the ResNet and MPS at different times during the post-quench dynamics.} \label{fig:spectra_app}
\end{figure}

Fig.~\ref{fig:spectra_app} shows the normalized spectra for both the NQS and MPS architectures at three different points in time throughout the post-quench time evolution, making qualitative differences between the two ansätze apparent: While the NQS spectrum smoothly varies from its maximum to minimum value, the MPS spectrum exhibits a clear hierarchical structure of eigenvalues. 
These drop smoothly from the maximum value to around $10^{-14}$, i.e. machine precision, where a large fraction of the spectrum resides.  
This part corresponds to the exact zeros of the MPS representation, i.e. its gauge-redundant variational directions.
This separation becomes even more pronounced during the time evolution, with even steeper drops towards machine precision.
For NQS, the spectrum remains qualitatively the same throughout the dynamics.

\section{Additional data regarding the principal angle analysis} \label{app:nullspace_rot}

\begin{figure}[h!]
    \includegraphics[width=\columnwidth]{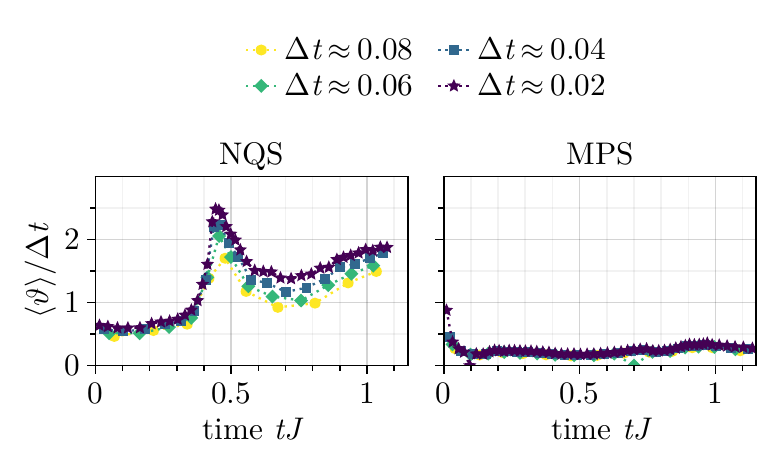}
    \caption{Mean principal angle normalized by the time separation, $\langle\vartheta\rangle/\Delta t$, for the NQS (left) and MPS (right). Results are shown for approximately $\Delta t=0.02$, $0.04$, $0.06$, and $0.08$, using points selected from the corresponding adaptive-step trajectories.} \label{fig:nullspace_rot_app}
\end{figure}

In this section, we investigate the dependence of the mean null-space rotation angle normalized by the time step, $\langle\vartheta\rangle/\Delta t$, on the chosen time step. 
The baseline is given by an adaptive-step time evolution performed for both the MPS and the ResNet. 
From that grid of points in time, we choose ones which are at least $\Delta t$ apart, and investigate the rotation angles for different choices of the timestep, see Fig.~\ref{fig:nullspace_rot_app}.
This allows us to check the robustness of the observed behavior.
We observe a mild dependence on the time step for NQS at $tJ>0.5$, with the curves appearing to converge as $\Delta t\rightarrow0$. 
For MPS, the results seem very stable across different choices of $\Delta t$.
This confirms that the observation is not an artifact of a finite time step but signals a genuine variation of the physically dominant and weakly relevant directions along the variational trajectory.
\bibliography{refs}
\end{document}